# Influencia del *Buffer* del *Router* en la Distribución de Vídeo P2P-TV

Idelkys Quintana-Ramírez, José Mª Saldaña, José Ruiz-Mas, Julián Fernández-Navajas, Luis Casadesus, Luis Sequeira

{idelkysq, jsaldana, jruiz, navajas, luis.casadesus, sequeira}@unizar.es

Grupo de Tecnologías de las Comunicaciones – Instituto de Investigación en Ingeniería de Aragón

Dpto. IEC. Centro Politécnico Superior Universidad de Zaragoza

Edif. Ada Byron, 50018, Zaragoza

*Abstract-* **This work presents a study of the behaviour of the router buffer when managing the traffic of P2P-TV applications, where a number of *peer*s exchange video content. First, a summary of the characteristics of SOPCast is presented. Then, the results obtained in simulation tests using different buffer policies are presented. Real traces of the application, obtained from a research project, have been used for the tests, sharing the Internet access with different amounts of background traffic. The results show that a similar buffer behaviour for all the access technologies. In addition, the big amount of small packets generated may impair the video traffic, thus avoiding the retransmission of the contents by the application.**

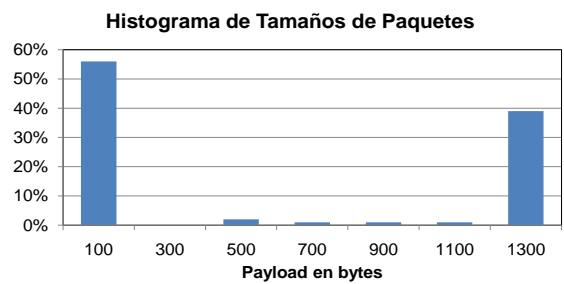

Fig. 1. Histograma del tamaño de los paquetes capturados.

## I. INTRODUCCIÓN

Actualmente se puede constatar el crecimiento de la distribución de contenidos de vídeo *online* y de servicios de televisión por Internet. Tradicionalmente, la solución utilizada para el *streaming* de vídeo por Internet ha sido el modelo cliente/servidor con variaciones como las denominadas *Content Delivery Network* (CDN) para disminuir la carga sobre un único servidor, aunque sigue siendo la escalabilidad su mayor reto. En este sentido el modelo *Peer-to-Peer* (P2P) surge como un nuevo paradigma en la distribución de contenidos de vídeo y TV. Es una solución práctica a los problemas de escalabilidad, al minimizar los costes requeridos tanto en equipos como en ancho de banda. Su rápida expansión muestra que las tecnologías orientadas a los usuarios y basadas en la colaboración entre ellos sin un control centralizado, son capaces de distribuir contenidos multimedia sensibles al retardo. Ejemplo de ello son aplicaciones como PPSTream, SOPCast, TVUPlayer y TVAnts que a día de hoy podemos encontrar en Internet.

La filosofía de diseño de este modelo es fomentar que los usuarios (*peer*) actúen como cliente y como servidor a la vez. En el modelo P2P, un *peer* no sólo descarga contenidos de la red sino que también participa en la distribución de los contenidos a otros *peer*. El sistema de *streaming* P2P se puede clasificar en dos categorías: *tree-based* y *mesh-based*. El más utilizado es el segundo, que posibilita que cada *peer* en la estructura de distribución actúe a la vez como consumidor y proveedor de los contenidos. Un *peer* puede tener múltiples *padres* (proveedores) y múltiples *hijos* (consumidores), lo cual supone un tráfico simultáneo tanto de subida como de bajada.

En la literatura pueden encontrarse diversos estudios y comparativas sobre este tipo de tráfico P2P. En [1] se presentó una caracterización del tráfico de las principales aplicaciones P2P-TV, concluyéndose que para cada una de ellas hay un patrón de tráfico diferente pero todas tienen en común el hecho de generar paquetes de gran tamaño (de vídeo) y paquetes pequeños (de señalización y de confirmación). En este sentido, la Fig. 1 muestra un histograma del tamaño de los paquetes, mayoritariamente UDP, generados por SOPCast y obtenidos a partir de las trazas recopiladas en [2]. Se observa que los tamaños se agrupan en torno a dos valores: el primero alrededor de los 100 *bytes* (aprox. 60%) y el segundo en los 1.320 *bytes* (aprox. 40%) que corresponden a los paquetes de vídeo. Los paquetes pequeños que corresponden a control de la capa de aplicación [3] son necesarios para gestionar del estado de los *peer* y poder reorganizar los paquetes aportados por todos ellos para una correcta reconstrucción del vídeo en el destino.

Los equipos de acceso a la red utilizados están diseñados para gestionar paquetes de gran tamaño [4], típicos de servicios tradicionales como el *e-mail*, la web o FTP. Sin embargo, vemos que los programas P2P-TV generan paquetes pequeños de señalización con una tasa lo suficientemente grande como para generar problemas de gestión en el *router* de acceso. Dicho *router* podría convertirse en un cuello de botella si se reciben demasiados paquetes por segundo.

En este trabajo estudiaremos el efecto que puede tener el comportamiento del *buffer* del *router* frente a diferentes tráficos de *streaming* de vídeo generado por SOPCast, teniendo en cuenta el tamaño medio de los paquetes y las diferentes tecnologías de acceso a Internet. La siguiente sección presenta los trabajos relacionados. La sección III resume el comportamiento del SOPCast. La sección IV presenta las pruebas y resultados y el trabajo se cierra con las conclusiones.

## II. TRABAJOS RELACIONADOS

El escenario que estamos considerando puede estar basado en diferentes tecnologías y lo mismo ocurre con el router de acceso puesto que hay una gran variedad. En los últimos años se han publicado muchos estudios sobre la problemática del dimensionado del *buffer* [5] aunque se ha abordado principalmente para los *router* del núcleo de la red y para flujos TCP. El problema de dimensionar el *buffer* fue estudiado en [6], donde se explica que hasta hace unos años se aceptaba la regla no escrita de usar el producto retardo-ancho de banda para dimensionarlo; pero esta regla está siendo sustituida por el llamado *"Stanford model"*, que propone *buffer* más pequeños. Posteriormente [7] se han propuesto tamaños menores (denominados *tiny buffer*) con una capacidad de varias decenas de paquetes. En este trabajo compararemos estas propuestas con los *buffer* grandes.

En la literatura existente se pueden encontrar distintas propuestas respecto a la influencia de los *buffer* en diferentes servicios. En [8], [9] y [10] se muestran resultados en los que diferentes comportamientos, tamaños y políticas de los *buffer* modifican la calidad de los servicios. El conocimiento de los dispositivos de acceso a la red permite tomar decisiones en cuanto a qué técnicas de optimización del tráfico aplicar. Ejemplo de ello son las propuestas sobre servicios en tiempo real como VoIP o juegos *online* [11], cuyo tráfico de paquetes pequeños es optimizado mediante técnicas de multiplexión [12] para ganar en eficiencia y ahorrar ancho de banda pero teniendo en cuenta que el aumento del tamaño de los paquetes puede a su vez perjudicar a la calidad para ciertas políticas de *buffer*.

En cuanto a las soluciones P2P existentes hay muchos y variados estudios sobre el impacto de su tráfico en las redes de comunicaciones [1][13][14][15], mejoras que aportan en las redes IPTV de operadores [16], percepción del usuario del servicio que proporcionan [17][18][19], alternativas en algoritmos de distribución [20], etc. Todas ellas abordan el análisis y mejora de los servicios P2P sin tener en cuenta que las ventajas de colaboración entre usuarios como el ahorro de ancho de banda y costes asociados al servicio pueden ser cercenadas por los propios dispositivos de acceso del usuario y sean sólo unos pocos usuarios *"privilegiados"* de la red P2P los que participen en la distribución de contenidos.

### III. RESUMEN DEL FUNCIONAMIENTO DE SOPCAST

SOPCast es una aplicación gratuita P2P utilizada para la transmisión de los programas de TV con tasas que varían en el rango de 250 kbps a 400 kbps, llegando a alcanzar en vídeos de alta calidad hasta 800 kbps.

Al inicio de la aplicación se requiere de un período de tiempo para realizar una búsqueda de *peer* activos en la estructura P2P de los que descargar datos. De este modo, al examinar el tráfico generado por cada nodo se observa que en primer lugar se envía un mensaje al servidor central de canales SOPCast (identificado por una IP), solicitando la obtención y actualización de una lista de canales.

Después de seleccionar el canal, el *peer* envía múltiples mensajes de solicitud a algunos servidores SOPCast para obtener una lista de *peer* activos que tengan sintonizado el canal deseado. Los *peer* se identifican por su dirección IP y número de puerto. Una vez recibido el listado de *peer*, el cliente envía mensajes para averiguar cuáles se encuentran activos. Estos pueden a su vez devolver sus propias listas de *peer*, ayudando al nuevo cliente a encontrar más *peer*.

En la Tabla 1 se resumen los principales paquetes que son intercambiados por la aplicación SOPCast, mostrando su tamaño y su funcionalidad. Los paquetes pequeños son de control y la descripción de su funcionalidad indica el propio funcionamiento del protocolo utilizado por SOPCast.

### IV. PRUEBAS Y RESULTADOS

El objetivo de las pruebas que se presentan es el estudio de la influencia mutua entre las políticas del *buffer* y las técnicas de distribución de tráfico P2P-TV. La generación de gran cantidad de paquetes pequeños [3] puede provocar que el *buffer* afecte a los paquetes de vídeo, de mayor tamaño, y también que el comportamiento de los *peer* en la estructura de la red P2P no sea el esperado. La política del *buffer* puede hacer que las pérdidas dependan del tamaño de paquete.

Las trazas de tráfico utilizadas se han obtenido del proyecto NAPA-WINE [2]. Las capturas se obtuvieron con distintas aplicaciones P2P-TV, entre ellas SOPCast, accediendo a Internet desde una LAN o una red doméstica del tipo ADSL. Se utilizaron 44 *peer* distribuidos en diferentes localizaciones geográficas de 4 países. La duración de los experimentos fue de una hora, donde todos los *peer* tenían seleccionado el mismo canal de TV. En todos los casos la velocidad de transmisión del vídeo fue de 348 kbps, y la calidad del vídeo percibida por los usuarios fue también similar. En las pruebas se utilizaran dos trazas: una obtenida con un ordenador conectado vía LAN a Internet (*High BW*) y otra usando ADSL.

La Fig.2 muestra el escenario empleado en las pruebas que se presentan. Se ha utilizado MATLAB para realizar simulaciones en las que el tráfico P2P y un tráfico de fondo comparten el ancho de banda del acceso. Las trazas se envían siguiendo exactamente los tiempos y tamaños originales, que se leen de sendos ficheros. Por tanto, no se ha necesitado modelar el tráfico. El tráfico de fondo tiene la siguiente distribución: el 50% son de 40 *bytes*, el 10% de 576, y el 40% restante de 1.500 [21].

Se han usado tres *buffer* con tamaño definido en *bytes* (10, 100 y 1000 *kbytes*) y otros tres limitados en número de paquetes (27, 270 y 2700 paq). Los tamaños son equivalentes considerando un tamaño medio de paquete de 370 *bytes*. Los ancho de banda de subida empleados son 512, 1024 y 2048 kbps, valores frecuentes en accesos a Internet actuales.

| Resumen del Tráfico SOPCast | | |
|---|---|---|
| Tipo de paquetes | Tamaño (bytes) | Funcionalidad |
| Vídeo | 1320 | Tamaño máximo de los paquetes de vídeo |
| | 377, 497, 617, 1081, 1201 | Fragmentos de los paquetes de vídeo |
| Control | 52 | Paquetes HELLO para iniciar las conexiones |
| | 80 | Confirmación de la recepción del paquete HELLO |
| | 28 | Paquetes de Confirmación (ACK) |
| | 42 | Mensaje de *Keep-alive* |
| | 46 | Paquetes de solicitud de vídeo |

Tabla 1. Resumen del Tráfico generado por SOPCast

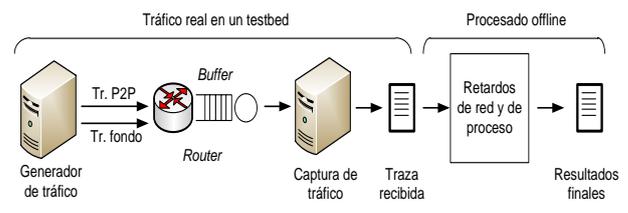

Fig. 2. Esquema de las pruebas.

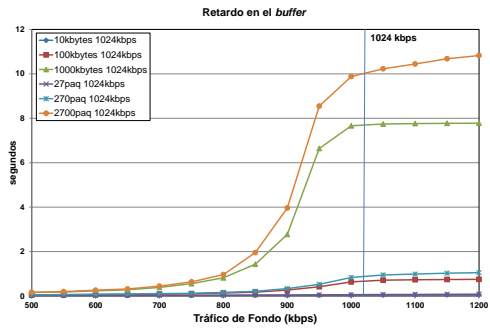

Fig. 3. Retardo para el ancho de banda 1024 kbps y High BW.

Se presentan algunas gráficas del retardo y pérdidas para las distintas políticas de *buffer* empleadas usando diferentes valores de tráfico de fondo para saturar el *router*. La Fig. 3 muestra el retardo usando un ancho de banda de 1024 kbps y la traza *High BW*. Vemos que el retardo crece conforme se alcanza el límite del ancho de banda y los *buffer* tienden a llenarse. Esta figura ilustra que el comportamiento es muy similar para ambos *buffer* siendo ligeramente mayor para el limitado en paquetes. Lo mismo ocurre para 512 y 2048 kbps y cuando se utiliza la traza de ADSL.

En la Fig. 4 se muestran las pérdidas de paquetes para un *peer* usando la traza *High BW*, para los tres posibles valores de ancho de banda. Vemos que las pérdidas en los *buffer* limitados en número de paquetes están muy por encima de las de los limitados en tamaño. Esto se debe a la gran cantidad de paquetes por segundo que genera la aplicación y que ocupan un lugar en el *buffer* sea cual sea su tamaño. La Fig. 5 es similar a la anterior pero utilizando la traza ADSL. Las gráficas vuelven a mostrar lo que se había planteado: en los *buffer* limitados en tamaño se tienen pérdidas mucho menores que en los limitados en número de paquetes.

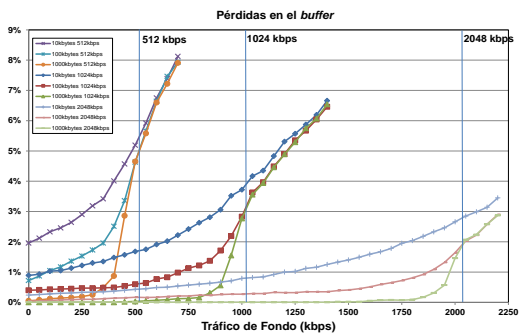

(a)

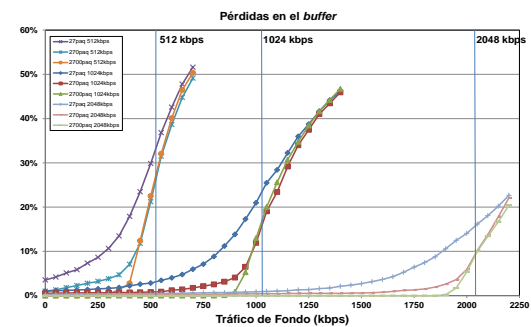

(b)

Fig. 4. Pérdidas para el *buffer* con la traza High BW: (a) limitado en tamaño, (b) limitado en número de paquetes.

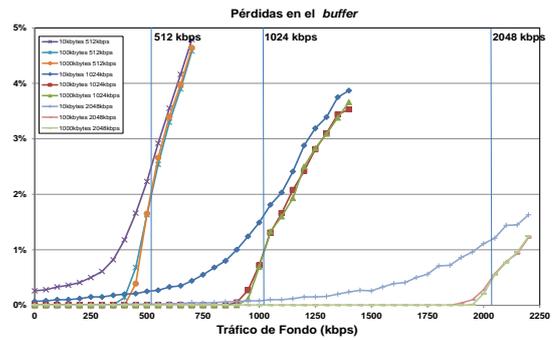

(a)

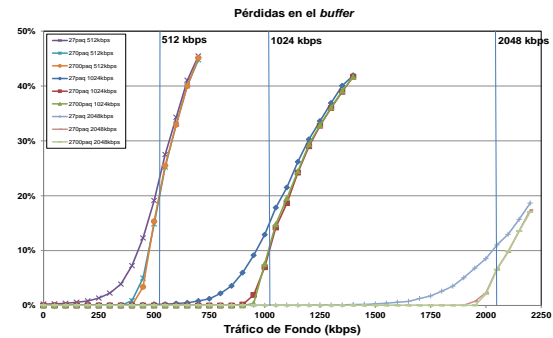

(b)

Fig. 5. Pérdidas para el *buffer* con la traza ADSL: (a) limitado en tamaño, (b) limitado en número de paquetes.

En la Fig.6 se presenta las pérdidas en el *buffer* según el tipo de paquetes. Estas figuras se muestran solo para un ancho de banda de salida de 1024 kbps pues como se ha mencionado el comportamiento es muy similar para 512 y 2048 kbps. Para los *buffer* medidos en *bytes* la pérdida de paquetes pequeños es mínima si se compara con la de los paquetes de vídeo, que son de tamaño mucho mayor. Como consecuencia de esto, puede ocurrir que el *peer* no contribuya suficientemente en la distribución de vídeo a otros usuarios. En cambio, para los *buffer* limitados en número de paquetes se desechan en la misma medida tanto los paquetes de vídeo como los de señalización. A pesar de su pequeño tamaño, cada paquete de señalización ocupa un puesto en el *buffer*. Esto hace que la cola se llene más rápido, penalizando a todos los paquetes y perjudicando tanto a la distribución como a la propia recepción de vídeo.

El efecto descrito es independiente de la tecnología de acceso empleada. En las redes *High BW* al incrementarse el ancho de banda de salida, los *peer* podrían enviar mayor número de paquetes de vídeo contribuyendo en mayor medida a la distribución. Sin embargo como se puede apreciar en la Fig. 6 las pérdidas pueden llegar a ser considerables si el *buffer* no es lo suficientemente grande.

Es más, el tráfico de fondo de paquetes grandes que se está desechando junto al tráfico de paquetes de vídeo, podría corresponder a otro tipo de servicios como FTP y HTTP. El buen funcionamiento de estos otros servicios también se vería perjudicado. Por otro lado, la utilización en el mismo enlace de servicios que utilicen paquetes pequeños, como VoIP y juegos *online* podrían llegar a perjudicar la distribución de vídeo P2P.

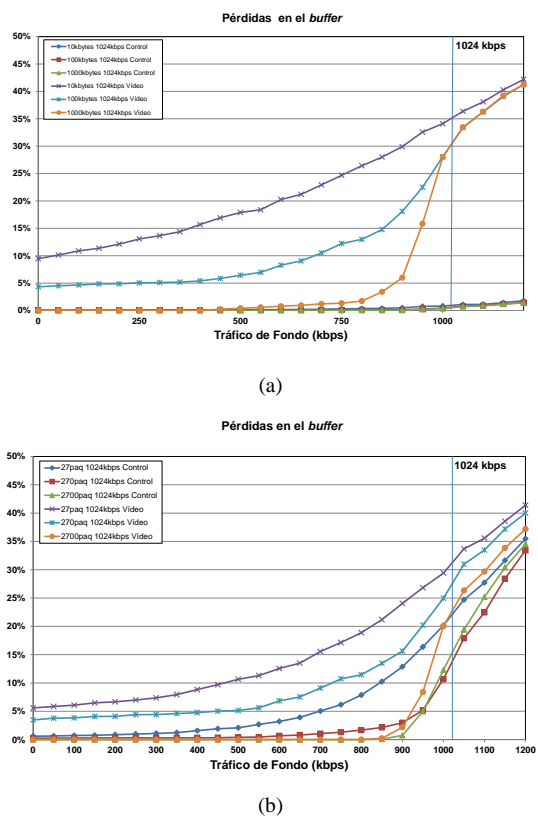

Fig. 6. Pérdidas según el tipo de paquetes con la traza *High BW* y 1024kbps: (a) limitado en tamaño, (b) limitado en número de paquetes

## V. Conclusiones

En este trabajo se ha presentado una comparativa del comportamiento del tráfico de una aplicación P2P-TV, en función de las políticas de *buffer*. Se ha mostrado que el *buffer* puede influir de manera negativa en el tratamiento dado a los paquetes de vídeo, provocando que los *peer* no contribuyan a la distribución de paquetes de vídeo. Se hace necesario por tanto tener en cuenta el tamaño de los paquetes que se generan, ya que algunas políticas penalizan los paquetes grandes.

Teniendo en cuenta la gran variedad de tecnologías de acceso que se pueden encontrar en el escenario considerado, las medidas presentadas ilustran que independientemente del ancho de banda de salida a Internet, el comportamiento para cada tipo de *buffer* es similar en cuanto a retardo. Sin embargo, se observa que las pérdidas obtenidas en los *buffer* limitados en *bytes* son mucho menores que en los limitados en número de paquetes.

Este trabajo forma parte de uno más amplio en el que se abordan temas tales como optimizar la distribución del tráfico P2P-TV, con el propósito de ahorrar ancho de banda y utilizar mas eficientemente los recursos de la red. Además, se pretende proporcionar a los fabricantes una técnica flexible que pueda ser útil para mejorar la experiencia del usuario en escenarios en que muchos clientes soliciten el mismo canal de TV.

## Referencias

[1] T. Silverton y O. Fourmaux, *"Measuring P2P IPTV Systems"*, en 17th International workshop on Network and Operating Systems Support for Digital Audio & Video, Proceedings of NOSSDAV'07, June 2007

[2] *"Network-Aware P2P-TV Application over Wise Networks"*, http://www.napa-wine.eu.

[3] B. Fallica, Yue Lu, F. Kuipers, R. Kooji & P. Van Mieghem, *"On the quality of experience of SopCast"*, NGMAST '08, pp. 501-506, September 2008.

[4] J.M. Saldaña, J. Fernández-Navajas, J. Ruiz Más, E. Viruete Navarro, L. Casadesus, *"The Utility of Characterizing Packet Loss as a Function of Packet Size in Commercial Routers"*, CCNC2012, pp. 362-363, January 2012

[5] A. Vishwanath, V. Sivaraman, and M. Thottan. *"Perspectives on router buffer sizing: recent results and open problems"*, SIGCOMM Comput. Commun. Rev. 39, 2, pp.34-39, March 2009.

[6] A. Dhamdhere and C. Dovrolis, *"Open issues in router buffer sizing"* Comput. Commun. Rev., vol. 36, no. 1, pp. 87-92, January 2006.

[7] M. Enachescu, Y. Ganjali, A. Goel, N. McKeown, T. Roughgarden. *"Part III: routers with very small buffers"*. SIGCOMM Comput. Commun. Rev. 35, 3, pp. 83-90, July 2005.

[8] J. Saldana, J. Fernandez-Navajas, J. Ruiz-Mas, E. Viruete Navarro y L. Casadesus, *"Influence of online games traffic multiplexing and router buffer on subjective quality"* in Proc. CCNC 2012- 4th IEEE International Workshop on Digital Entertainment, Networked Virtual Environments, and Creative Technology (DENVECT), pp. 482–486, Las Vegas, January 2012.

[9] J. Saldana, J. Murillo, J. Fernandez-Navajas, J. Ruiz-Mas, E. Viruete, and J. I. Aznar, *"Evaluation of multiplexing and buffer policies influence on VoIP conversation quality"*, In Proc. CCNC 2011- 3rd IEEE International Workshop on Digital Entertainment, Networked Virtual Environments, and Creative Technology, pp. 1147–1151, Las Vegas, January 2011

[10] J.M. Saldaña, J. Fernández-Navajas, J. Ruiz Más, J. I. Aznar, E. Viruete Navarro, L. Casadesus, *"Influencia del buffer del router en la multiplexión de juegos online"*, URSI 2011, Septiembre 2011.

[11] J. Saldana, J. Murillo, J. Fernández-Navajas, J. Ruiz-Mas, J. I. Aznar, E. Viruete, *"Bandwidth efficiency improvement for online games by the use of tunneling, compressing and multiplexing techniques"*, unpublished.

[12] B. Thompson, T. Koren, D. Wing. *"RFC 4170: Tunneling Multiplexed Compressed RTP"*, November 2005.

[13] T. Silverston, O. Fourmaux, K. Salamatian, K. Cho, *"Measuring P2P-TV systems on both sides of the world"*, in: 2nd International Workshop on IPTV Technologies and Multidisciplinary Applications (IWITMA), IEEE ICME, 2010.

[14] S. Tang, Y. Lu, J.M. Hernández, F.A. Kuipers, and P. Van Mieghem. *"Topology dynamics in a P2PTV network"*. In Proc. Networking 2009, 326-337, 2009.

[15] Eittenberger, P., Krieger, U.R., Markovich, N.M.: *"Measurement and Analysis of Live-Streamed P2PTV Traffic"*. In: T. Czachórski (ed.), Performance Modelling and Evaluation of Heterogeneous Networks, Proc. HET-NETs 2010, Zakopane, Poland, January, 2010.

[16] M. Cha, P. Rodriguez, S. Moon, and J. Crowcroft, *"On next-generation telco-managed P2P TV architectures"*, in IPTPS '08, 2008

[17] Y. Lu; B. Fallica; F. A. Kuipers; R. E. Kooij & P. V. Mieghem, *"Assessing the Quality of Experience of SopCast"*, IJIPT 4 (1), pp. 11-23, 2009.

[18] Fallica, B., Lu, Y., Kuipers, F.A., Kooij, R.E. and Van Mieghem, P: *"On the Quality of Experience of SopCast"*, 1st IEEE International Workshop on Future Multimedia Networking FMN'08, Cardiff, Wales, UK, 2008

[19] U.R. Krieger, R. Schweßinger, *"Analysis and Quality Assessment of Peer-to-Peer IPTV Systems"*, In Proc. 12th Annual IEEE International Symposium on Consumer Electronics (ISCE2008), April 2008

[20] Y. Liu, Y. Guo, and C. Liang, *"A survey on peer-to-peer video streaming systems"*, Journal of *Peer*-to-*Peer* Networking and Applications, vol. 1, no. 1, pp. 18–28, March 2008.

[21] Cooperative Association for Internet Data Analysis: NASA Ames Internet Exchange Packet Length Distributions.